\documentclass[aps, pra, twocolumn, superscriptaddress, longbibliography]{revtex4-2}
\makeatletter
\@booleanfalse\footinbib@sw
\makeatother

\usepackage{amsmath}
\usepackage{amssymb}
\usepackage{bbm}
\usepackage{graphicx}
\usepackage{subcaption}
\usepackage{hyperref}
\usepackage{lipsum}
\usepackage{array}
\usepackage{booktabs}
\usepackage{tikz}
\usepackage{quantikz}
\usepackage{xcolor}
\usepackage{tikz-3dplot}
\usepackage{acronym}
\usepackage[capitalise]{cleveref}

\usepackage{orcidlink}

\newacro{TN}[TN]{tensor network}
\newacroplural{TN}[TNs]{tensor networks}
\newacro{RO}[RO]{Riemannian optimization}
\newacro{PO}[PO]{hardware-aware parameterized optimization}
\newacro{TLFIM}[TLFIM]{transverse-longitudinal-field Ising model}
\newacro{MPS}[MPS]{matrix product state}
\newacroplural{MPS}[MPS]{matrix product states}
\newacro{MPO}[MPO]{matrix product operator}
\newacroplural{MPO}[MPOs]{matrix product operators}
\newacro{SVD}[SVD]{singular-value decomposition}
\newacro{DOF}[DOF]{degrees of freedom}
\newacro{NISQ}[NISQ]{noisy intermediate-scale quantum}

\definecolor{evenColor}{RGB}{173, 216, 230}
\definecolor{oddColor}{RGB}{255, 182, 193}

\hypersetup{
    colorlinks=true,
    linkcolor=blue,
    citecolor=blue,
    urlcolor=blue
}

\begin{document}

\title{Exploiting the Weyl-Chamber Structure for Hardware-Aware Quantum Circuit Compression}

\author{Neel Misciasci\,\orcidlink{0009-0000-0711-7023}}
\email{neel.misciasci@tum.de}
\affiliation{Volkswagen AG}
\affiliation{Technical University of Munich, School of Computation, Information and Technology, Department of Computer Science, Boltzmannstr.~3, 85748 Garching, Germany}

\author{Isabel~Nha~Minh~Le\,\orcidlink{0000-0001-6707-044X}}
\email{isabel.le@tum.de}
\affiliation{Technical University of Munich, School of Computation, Information and Technology, Department of Computer Science, Boltzmannstr.~3, 85748 Garching, Germany}
\affiliation{Munich Center for Quantum Science and Technology (MCQST), Schellingstr.~4, 80799 Munich, Germany}

\author{Christian~B.~Mendl\,\orcidlink{0000-0002-6386-0230}}
\email{christian.mendl@tum.de}
\affiliation{Technical University of Munich, School of Computation, Information and Technology, Department of Computer Science, Boltzmannstr.~3, 85748 Garching, Germany}
\affiliation{Munich Center for Quantum Science and Technology (MCQST), Schellingstr.~4, 80799 Munich, Germany}
\affiliation{Technical University of Munich, Institute for Advanced Study, Lichtenbergstr.~2a, 85748 Garching, Germany}

\date{\today} 

\begin{abstract}
Quantum circuit compression based on tensor networks enables long-time simulation of quantum many-body dynamics on near-term hardware.
While Riemannian optimization has been successfully employed for this purpose, it typically optimizes over the manifold of generic two-qubit unitaries.
Consequently, the resulting gates must still be synthesized into the hardware's native gate set via KAK decomposition -- an overhead that partially negates the original compression gain. 
We show that Riemannian-optimized gates already exhibit a predictable Weyl structure: near-$R_{ZZ}$ for the Ising model and near-fSim for the Heisenberg XXZ model. 
Building on this insight, we investigate whether parameterized optimization directly over the native gate set can offer a beneficial trade-off between expressivity and circuit efficiency by reducing the circuit's dimensionality.
When comparing the two approaches, we find that the small residual Weyl parameters neglected by the latter are essential for high compression of the Ising propagator, whereas for the Heisenberg propagator, the native XXZ structure can be exploited directly to build a more efficient compressed circuit.
\end{abstract}

\maketitle 

\section{Introduction}
\label{sec:intro}
Simulating the time evolution of quantum systems, particularly in quantum many-body physics, remains a primary motivation for developing quantum computers~\cite{lloyd1996universal}. 
In this work, we consider a time-independent Hamiltonian $H$ and aim to implement the time evolution operator $e^{-iHt}$. 
Besides the qubitization framework~\cite{low2019hamiltonian}, a natural procedure is using Trotterization-type circuit ansätze~\cite{childs2021trotter}, which do neither require block-encoding nor additional auxiliary qubits.

Several works have demonstrated that ``vanilla'' Trotterization circuits based on product formulas can be further optimized to improve the approximation accuracy and to reduce the circuit depth~\cite{kotil2024riemannianquantum,le2025riemannianquantum,zhang2024scalablequantumdynamicscompilation,gibbs2025deepcircuit,gibbs2025learning, guo2025efficientquantum,dinca2025quantumprocesstomographycompressed, mansuroglu_parametric,mc_keever_parametric,robertson_parametric}.
One prominent pathway combines \ac{RO} with \acp{TN}~\cite{kotil2024riemannianquantum,le2025riemannianquantum,putterer2025high,karacan2026nearoptimalquantumtimeevolution,le2026hessianvectorproductstensornetworks} to reduce the number of layers in a Trotterized circuit by optimizing over the manifold of generic unitaries. 
However, a significant practical bottleneck emerges during hardware execution: \ac{RO} returns dense $4\times4$ unitaries, each of which must be decomposed into the underlying native gate set.
This synthesis naturally increases circuit depth, partially undoing the initial compression~\cite{hung2025improvedisingmesonspectroscopy}.
The Cartan coordinates of these gates characterize their non-local entangling content and reveal predictable, model-dependent structures, suggesting that some systems may not require the expressivity of the full unitary manifold.
This raises the possibility that a lower-dimensional native ansatz can retain the relevant expressivity while reducing synthesis overhead.
We therefore ask: can optimization restricted to a hardware-native gate family achieve lower approximation error at a fixed native entangling depth than optimization over generic two-qubit unitaries followed by synthesis?

Motivated by these structural patterns, we suggest an alternative \ac{PO} method and compare it with \ac{RO}.
While optimizing structured, parameterized ansätze has been explored in classical compression schemes~\cite{mansuroglu_parametric, mc_keever_parametric, robertson_parametric}, our approach expands on these foundations in three key ways. 
First, rather than relying on explicit translational invariance to extrapolate from small chains~\cite{mansuroglu_parametric}, we optimize the full $N=50$ chain directly through \ac{MPO} gradients.
Second, we explicitly address the expressivity limits of native gate sets -- a question previously left open~\cite{mc_keever_parametric}. 
Finally, we generate a reusable time-evolution operator capable of reaching longer times, in contrast to methods that compile single-target state-preparation circuits~\cite{robertson_parametric}.

When comparing \ac{PO} and \ac{RO}, we observe that \ac{PO} struggles to match \ac{RO}'s generalizability and low loss for the Ising propagator. In contrast, the local XXZ structure in the Heisenberg propagator can be leveraged to construct a superior compressed circuit. 
These results inform the design of more resource-efficient hardware experiments~\cite{hung2025improvedisingmesonspectroscopy,chai2025resource}.

\section{Motivation}
\label{sec:motivation}
In order to execute a quantum circuit experimentally, its constituent quantum gates ultimately need to be decomposed into the native gate set of the target hardware~\cite{PhysRevA.69.032315,vidal_2004_universal_quantum_circuit,PhysRevA.69.062321}. 
To assess how compression using \ac{TN} methods affects the two-qubit gate count and circuit depth after decomposition into the target hardware's native gate set, we first examine the fundamental algebraic structure of two-qubit gates through the KAK decomposition -- the Cartan decomposition of the $SU(4)$ Lie group~\cite{kraus2001optimalcreationofentanglement,KHANEJA200111}. 
This theorem states that any arbitrary two-qubit unitary $U \in SU(4)$ can be factored into a canonical, multi-parameter entangling core flanked by local single-qubit rotations:
\begin{equation}
\begin{split}
    U &= (K_{1} \otimes K_{2}) \\
    &\quad\times \exp \left[ -i \alpha_x X \otimes X + \alpha_y Y \otimes Y + \alpha_z Z \otimes Z) \right] \\
    &\quad\times (K_{3} \otimes K_{4})
\end{split}
\label{eq:kak}
\end{equation}
\[
\begin{quantikz}[align equals at=1.5]
    & \gate[wires=2]{U} & \qw 
    \midstick[2,brackets=none]{=} 
    & \gate{K_3} & \gate[wires=2][1.5cm]{e^{-i\vec{\alpha}\cdot\vec{\sigma}\otimes\vec{\sigma}}} & \gate{K_1} & \qw \\
    &                   & \qw 
    & \gate{K_4} & \ghost{e^{-i\vec{t}\cdot\vec{\sigma}\otimes\vec{\sigma}}}               & \gate{K_2} & \qw
\end{quantikz}
\]
Here, $K_1, K_2, K_3, K_4 \in SU(2)$ are local single-qubit rotations, and $X, Y, Z$ are the standard Pauli matrices.
The real-valued parameters $(\alpha_x, \alpha_y, \alpha_z)$ are the so-called \emph{Weyl} or \emph{Cartan coordinates}~\cite{zhang2004minimum,wu2025designquantuminstructionset}, which fully characterize the non-local, entanglement of $U$ up to local equivalence~\cite{zhang2004minimum,makhlin2002nonlocal}.

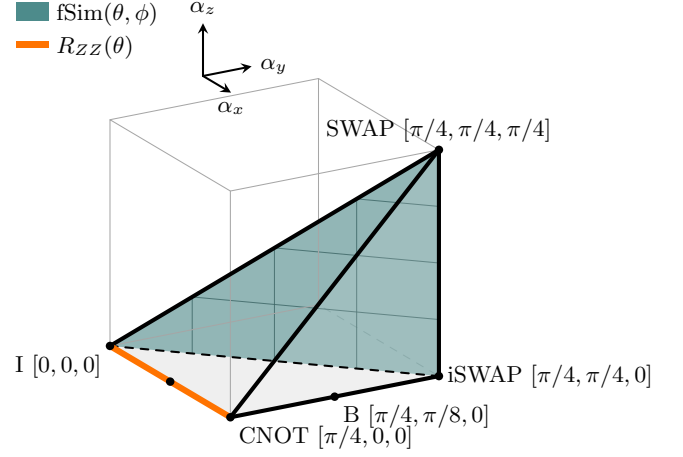
\begin{figure}[htbp]
    \centering
    \tdplotsetmaincoords{70}{60}

    \resizebox{\columnwidth}{!}{%
    \begin{tikzpicture}[tdplot_main_coords, scale=1.1, line join=round, line cap=round, >=stealth]

        \def\QPI{3.0}
        \def\EPI{1.5}

        \coordinate (O)  at (0, 0, 0);
        \coordinate (A1) at (\QPI, 0, 0);
        \coordinate (A2) at (\QPI, \QPI, 0);
        \coordinate (A3) at (\QPI, \QPI, \QPI);
        \coordinate (L)  at (\EPI, 0, 0);
        \coordinate (B)  at (\QPI, \EPI, 0);

        \draw[gray!50, dashed, thin] (\QPI, 0, 0) -- (\QPI, \QPI, 0) -- (0, \QPI, 0);
        \draw[gray!50, dashed, thin] (\QPI, \QPI, 0) -- (\QPI, \QPI, \QPI);

        \colorlet{fsimColor}{teal!70!black}
        \colorlet{rzzColor}{orange!90!red}
        \fill[gray!30, opacity=0.4] (O) -- (A1) -- (A2) -- cycle;
        \fill[fsimColor, opacity=0.55] (O) -- (A2) -- (A3) -- cycle;
        \foreach \t in {0.25, 0.5, 0.75} {
            \draw[fsimColor!60!black, thin, opacity=0.8] (\t*\QPI, \t*\QPI, \t*\QPI) -- (\QPI, \QPI, \t*\QPI);
            \draw[fsimColor!60!black, thin, opacity=0.8] (\t*\QPI, \t*\QPI, 0) -- (\t*\QPI, \t*\QPI, \t*\QPI);
        }
        \fill[gray!10, opacity=0.25] (O) -- (A1) -- (A3) -- cycle;
        \fill[gray!10, opacity=0.35] (A1) -- (A2) -- (A3) -- cycle;

        \draw[gray!70, thin] (0,0,0) -- (0,0,\QPI) -- (\QPI,0,\QPI) -- (\QPI,0,0);
        \draw[gray!70, thin] (0,0,\QPI) -- (0,\QPI,\QPI) -- (\QPI,\QPI,\QPI) -- (\QPI,0,\QPI);
        \draw[gray!70, thin] (0,\QPI,\QPI) -- (0,\QPI,0) -- (0,0,0);

        \draw[black, dashed, thick] (O) -- (A2);
        
        \draw[black, ultra thick] (O) -- (A1);
        \draw[black, ultra thick] (A1) -- (A2);
        \draw[black, ultra thick] (A2) -- (A3);
        \draw[black, ultra thick] (A3) -- (O);
        \draw[black, ultra thick] (A1) -- (A3);
        \draw[rzzColor, line width=2.4pt] (O) -- (A1);

        \foreach \p in {O, A1, A2, A3, L, B} {
            \fill[black] (\p) circle (1.5pt);
        }

        \node[anchor=north east] at (O)  {\small I $[0, 0, 0]$};
        \node[anchor=north west] at (A1) {\small CNOT $[\pi/4, 0, 0]$};
        \node[anchor=west]       at (A2) {\small iSWAP $[\pi/4, \pi/4, 0]$};
        \node[anchor=south]      at (A3) {\small SWAP $[\pi/4, \pi/4, \pi/4]$};
        \node[anchor=north west] at (B)  {\small B $[\pi/4, \pi/8, 0]$};

        \begin{scope}[shift={(-0.8, 1.8, \QPI)}]
            \draw[->, thick] (0,0,0) -- (0.7,0,0) node[anchor=north] {$\alpha_x$};
            \draw[->, thick] (0,0,0) -- (0,0.7,0) node[anchor=west] {$\alpha_y$};
            \draw[->, thick] (0,0,0) -- (0,0,0.7) node[anchor=south] {$\alpha_z$};
        \end{scope}

        \node[anchor=north west, align=left, font=\small] at (current bounding box.north west) {%
            \textcolor{fsimColor!70}{\rule{0.45cm}{0.25cm}}~$\mathrm{fSim}(\theta,\phi)$\\[2pt]
            \textcolor{rzzColor}{\rule[0.5ex]{0.45cm}{2.4pt}}~$R_{ZZ}(\theta)$};

    \end{tikzpicture}%
    }
    \caption{Geometric representation of the Weyl chamber for two-qubit local equivalence classes. The chamber forms a tetrahedron within the reduced bounding region $\pi/4 \ge \alpha_x \ge \alpha_y \ge \alpha_z \ge 0$. Distinct high-symmetry operating points correspond to core quantum gates: Identity ($\text{I}$), $\text{CNOT}$, $\text{iSWAP}$, and $\text{SWAP}$ define the tetrahedral vertices, while the $\text{B}$ gate is positioned strictly on the base edge directly between $\text{CNOT}$ and $\text{iSWAP}$. The native gate families used in this work are highlighted: $R_{ZZ}(\theta)$ gates lie on the I--CNOT edge (orange), whereas $\mathrm{fSim}(\theta,\phi)$ gates covers the face $\alpha_x = \alpha_y$ (I--iSWAP--SWAP)  (teal).}
    \label{fig:weyl_chamber}
\end{figure}

The local gauge freedom can be fixed by confining these coordinates to a canonical sub-volume known as the \emph{Weyl chamber}, defined by the ordering $\frac{\pi}{4} \geq \alpha_x \geq \alpha_y \geq \alpha_z \geq 0$ (Fig.~\ref{fig:weyl_chamber}). 
Points in the bulk of the chamber correspond to \emph{generic} two-qubit gates, with all three coordinates non-zero.
Synthesizing such a generic unitary on standard hardware requires, in the worst case, three maximally entangling two-qubit gates (e.g., CNOT, CZ, or an equivalent parametric rotation) interleaved with single-qubit rotations~\cite{vidal_2004_universal_quantum_circuit}.
By contrast, gates on the face $\alpha_z = 0$ of the chamber, such as the fractional iSWAP or the B gate~\cite{zhang2004minimum}, can be synthesized
with only two CNOT-equivalent entangling gates, and gates locally equivalent
to CNOT itself, $(\pi/4, 0, 0)$, require a single one~\cite{PhysRevA.69.062321}.
This distinction is central to our motivation: for a given native gate set, the entangling-gate cost of a circuit after decomposition is determined not only by the number of two-qubit gates it contains, but also by \emph{where in the Weyl chamber} those gates lie.

\cref{fig:trotter_comparison}a illustrates this structural overhead for a circuit composed of generic two-qubit unitaries.
To address this limitation, we aim to preserve the advantages of the \ac{TN} optimization framework developed in Ref.~\cite{le2025riemannianquantum} while adapting it to a hardware-native, parameterized ansatz.

\begin{figure*}[t]
    \centering
    \tikzset{
        trotEven/.style={fill=evenColor},
        trotOdd/.style={fill=oddColor},
        trotLoc/.style={fill=gray!15},
        trotOneQ/.style={fill=gray!15, minimum width=0.32cm, minimum height=0.32cm, inner sep=0pt},
        trotZZ/.style={minimum width=0.5cm, inner xsep=1pt},
    }
    \newcommand{\trotU}{\gate[style={trotOneQ}]{}}
    \newcommand{\trotZZe}{\gate[2, style={trotEven, trotZZ}]{\scriptstyle R_{ZZ}}}
    \newcommand{\trotZZo}{\gate[2, style={trotOdd, trotZZ}]{\scriptstyle R_{ZZ}}}

    {\raggedright\small\textbf{(a)}~Riemannian splitting [Eq.~\eqref{eq:trotter_split_riemannian}]\par}
    \vspace{0.3em}
    \begin{quantikz}[row sep=0.3cm, column sep=0.09cm]
        \lstick{$q_0$} & & \gate[2, style={trotEven}]{G_1^1}\gategroup[4, steps=3, style={dashed, rounded corners, draw=gray, inner xsep=3pt}, label style={label position=above, anchor=south, yshift=-2pt}]{\footnotesize generic $SU(4)$ gates} & & \gate[2, style={trotEven}]{G_1^3} & \midstick[4, brackets=none]{$\xrightarrow{\ \text{synthesis}\ }$} &
            \trotU\gategroup[4, steps=19, style={dashed, rounded corners, draw=gray, inner xsep=3pt}, label style={label position=above, anchor=south, yshift=-2pt}]{\footnotesize native gates: $R_{ZZ}$ + single-qubit rotations} & \trotZZe & \trotU & \trotZZe & \trotU & \trotZZe & \trotU & & & & & & \trotU & \trotZZe & \trotU & \trotZZe & \trotU & \trotZZe & \trotU & \\
        \lstick{$q_1$} & & & \gate[2, style={trotOdd}]{G_1^2} & & &
            \trotU & & \trotU & & \trotU & & \trotU & \trotZZo & \trotU & \trotZZo & \trotU & \trotZZo & \trotU & & \trotU & & \trotU & & \trotU & \\
        \lstick{$q_2$} & & \gate[2, style={trotEven}]{G_2^1} & & \gate[2, style={trotEven}]{G_2^3} & &
            \trotU & \trotZZe & \trotU & \trotZZe & \trotU & \trotZZe & \trotU & & \trotU & & \trotU & & \trotU & \trotZZe & \trotU & \trotZZe & \trotU & \trotZZe & \trotU & \\
        \lstick{$q_3$} & & & & & &
            \trotU & & \trotU & & \trotU & & \trotU & & & & & & \trotU & & \trotU & & \trotU & & \trotU &
    \end{quantikz}

    \vspace{1.5em}

    {\raggedright\small\textbf{(b)}~Hardware-aware splitting [Eq.~\eqref{eq:trotter_split_good_hardware_aware}]\par}
    \vspace{0.3em}
    \begin{quantikz}[row sep=0.3cm, column sep=0.35cm]
        \lstick{$q_0$} & \gate[2, style={trotEven}]{R_{ZZ}(\Delta t/2)} & & \gate[style={trotLoc}]{U_{\mathrm{loc}}(\Delta t)} & \gate[2, style={trotEven}]{R_{ZZ}(\Delta t/2)} & & \\
        \lstick{$q_1$} & & \gate[2, style={trotOdd}]{R_{ZZ}(\Delta t/2)} & \gate[style={trotLoc}]{U_{\mathrm{loc}}(\Delta t)} & & \gate[2, style={trotOdd}]{R_{ZZ}(\Delta t/2)} & \\
        \lstick{$q_2$} & \gate[2, style={trotEven}]{R_{ZZ}(\Delta t/2)} & & \gate[style={trotLoc}]{U_{\mathrm{loc}}(\Delta t)} & \gate[2, style={trotEven}]{R_{ZZ}(\Delta t/2)} & & \\
        \lstick{$q_3$} & & & \gate[style={trotLoc}]{U_{\mathrm{loc}}(\Delta t)} & & &
    \end{quantikz}

    \caption{One repetition of second-order Trotterization under the two splittings (Ising case). Blue and pink boxes act on even and odd bonds, respectively; grey boxes are single-qubit gates. \textbf{(a)} With the splitting used for Riemannian optimization [Eq.~\eqref{eq:trotter_split_riemannian}], each generic two-qubit gate $G \in SU(4)$ must be synthesized into the native gate set, here three $R_{ZZ}$ gates interleaved with single-qubit rotations; adjacent single-qubit rotations are merged. This triples the two-qubit gate depth. \textbf{(b)} With the hardware-aware splitting [Eq.~\eqref{eq:trotter_split_good_hardware_aware}], the Trotter step is implemented directly with native gates, yielding a shallower circuit.}
    \label{fig:trotter_comparison}
\end{figure*}

\begin{figure*}[t]
      \centering
      \begin{subfigure}[b]{\columnwidth}
          \includegraphics[width=\linewidth]{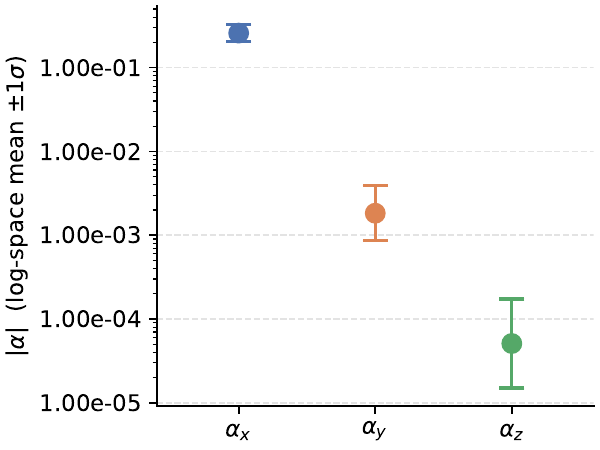}
          \caption{Ising model [Eq.~\eqref{eq:ising_hamiltonian}] with $J{=}1,\, h{=}0.75,\, g{=}0.6,\, t{=}2.0$.
           Coordinates are shown as log-space mean $\pm\,1\sigma$, since the three
           components span several orders of magnitude and a linear violin plot
           would obscure $\alpha_y$ and $\alpha_z$. The dominant coordinate
           $\alpha_x \gg \alpha_y, \alpha_z$ shows that nearly every optimized
           two-qubit gate collapses onto the first axis of the Weyl chamber,
           i.e.\ an $R_{ZZ}$-like gate, consistent with the pure $ZZ$ coupling of the Ising Hamiltonian.}
          \label{fig:weyl_ising}
      \end{subfigure}%
      \hfill
      \begin{subfigure}[b]{\columnwidth}
          \includegraphics[width=\linewidth]{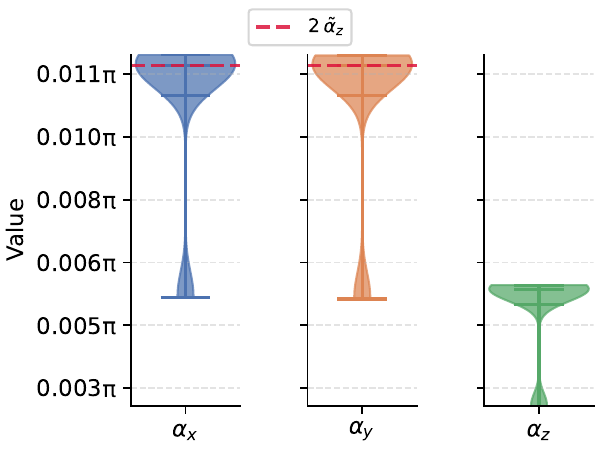}
            \caption{Heisenberg XXZ model [Eq.~\eqref{eq:heisenberg_hamiltonian}] with $J{=}1,\, J_z{=}{-}0.5,\, h{=}0.75,\, t{=}0.25$.
           Violins show the full distribution of each Weyl coordinate over all
           optimized two-qubit gates (mirrored about the central axis, per the
           usual violin-plot convention), with horizontal lines marking the mean,
           median, maximum, and minimum. The near-equality $\alpha_x \approx \alpha_y$
           reflects the U(1) symmetry of the XXZ interaction (equal $XX$ and $YY$ couplings $J$),
           while $\alpha_x \approx 2|\alpha_z|$ (dashed line, $2\tilde\alpha_z$)
           directly encodes the coupling ratio $J/|J_z| = 2$ between the $XY$ and $ZZ$
           terms of the Hamiltonian.}          
           \label{fig:weyl_heisenberg}
      \end{subfigure}
      \caption{Weyl-chamber coordinates $(\alpha_x, \alpha_y, \alpha_z)$ distribution of the \ac{RO}-optimized
           two-qubit gates for $N{=}50$ sites, extracted via the KAK
           decomposition, for two Trotterized target Hamiltonians. Both circuits
           use the same initial ansatz: 7 repetitions of a second-order Trotter
           product formula. The gate content reflects the symmetry of the
           underlying Hamiltonian: Ising optimization converges to near-$\mathrm{R}_{ZZ}$
           gates (a), while Heisenberg XXZ optimization converges to near-fSim gates (b).}
      \label{fig:weyl_parameters}
  \end{figure*}

Analyzing the Weyl coordinate distributions of \ac{RO}-optimized circuits reveals a striking structural pattern: despite having the entire generic manifold available for optimization, the resulting matrices naturally adopt simplified structures that mirror the local interactions of the simulated Hamiltonian:
\begin{itemize}
\item \textit{Ising model}: For each of the optimized gates, one Weyl coordinate is consistently several orders of magnitude larger than the other two (\cref{fig:weyl_ising}). Each optimized gate is therefore locally close to a single $R_{ZZ}$ rotation,\footnote{Strictly, the dominant coordinate is $\alpha_x$, i.e., an $XX$ rotation; $R_{XX}$ and $R_{ZZ}$ are related by single-qubit Hadamard gates and are thus locally equivalent. We choose $R_{ZZ}$ because it matches the interaction term of the model and is natively available on current hardware~\cite{egorova2025tunable, quantinuum_rzz}.} which requires one native entangling gate instead of the three needed for a generic $SU(4)$ gate. Most of the synthesis overhead in \cref{fig:trotter_comparison}a is thus spent on two small, apparently subleading coordinates; in \cref{sec:ising} we test whether they can be dropped by optimizing $R_{ZZ}$ gates directly.
\item \textit{Heisenberg XXZ model}:
A similar phenomenon occurs in the Heisenberg model. After optimization on the arbitrary $SU(4)$ manifold, the Weyl spectrum demonstrates that the gates strongly favor a local XXZ structure, $\alpha_x \approx \alpha_y$ (\cref{fig:weyl_heisenberg}). Such gates are locally equivalent to a single fSim gate~\cite{foxen_2020_demonstrating,jiang2025concurrent}, so in \cref{sec:heisenberg} we optimize parameterized fSim gates directly, rather than synthesizing each generic gate to an fSim and an $R_{ZZ}$ gate.
\end{itemize}
These observations suggest that, for these models, the optimized gates occupy only
a small subset of the full 15-parameter $SU(4)$ manifold. 
This motivates our core hypothesis: instead of optimizing over the full $SU(4)$ manifold and then paying the full cost of synthesizing each gate into native operations, we can optimize directly over native gates, with significantly fewer degrees of freedom, without a substantial loss in simulation accuracy. 
Such a hardware-native ansatz avoids the synthesis overhead altogether. 
Furthermore, because each hardware-native Trotter step is shallower than its synthesized $SU(4)$ counterpart (\cref{fig:trotter_comparison}b), more steps fit within a fixed native entangling-gate depth. 
The ansatz can therefore be initialized from a Trotter circuit with a smaller time step, and hence a lower Trotter error.

\section{Methods}
\label{sec:methods}

We compare two distinct \ac{TN}-based approaches for compressing 1D time-evolution circuits: standard \ac{RO} and \ac{PO}.
Both approaches rely on the same underlying computational engine, namely efficient \ac{TN} contraction to obtain loss gradients, but differ in the initial ansatz (as seen in \cref{fig:trotter_comparison}) over which the two-qubit gates are optimized.
Given a circuit $W$ composed of gates $G = \{G_i\}_{i \in \mathrm{ind}(W)}$, the goal in both approaches is to minimize the same loss $\mathcal{L}(U_{\text{ref}}, W)$ between $W$ and a reference time-evolution unitary $U_{\text{ref}}$.
We compare the resulting approximation error at fixed native entangling depth, including the synthesis of generic two-qubit gates in the \ac{RO} circuits.

\subsection{Riemannian optimization}
\label{subsec:riemannian_optimization}

As a baseline, we use the \ac{TN} \ac{RO} framework of Ref.~\cite{le2025riemannianquantum}, which optimizes generic two-qubit unitaries in a brickwall circuit directly on the unitary manifold. 
The reference propagator is represented as an \ac{MPO}, and the gate-matrix gradient $\partial\mathcal{L}/\partial G_i$ is obtained by contracting the environment of each gate $G_i$. 
This Euclidean gradient is projected onto the tangent space of $U(4)$ to give the Riemannian gradient, and the updated gate is mapped back onto the manifold by a retraction, so unitarity is preserved throughout. 
We use the Riemannian ADAM optimizer with a polar retraction~\cite{le2025riemannianquantum}. 
After optimization, each two-qubit gate is synthesized into the native gate set via its KAK decomposition (\cref{eq:kak}).

\subsection{Hardware-aware parameterized optimization}
\label{subsec:hw_optimization}

Justified by the structural behaviors observed in Sec.~\ref{sec:motivation}, \ac{PO} restricts the optimization to the native gate set of the target hardware instead of allowing for generic two-qubit unitaries.
To this end, the circuit is initialized in a Trotterization split natively aligned with that gate set. 
The optimization is performed using a hybrid scheme:
\begin{itemize}
    \item The native two-qubit gates are parameterized directly by their physical control parameters $\{\theta_i\}$ (e.g., the angles of an $R_{ZZ}$ or fSim gate) and updated via Euclidean gradient descent. The loss gradient with respect to these parameters is computed via the chain rule:
    \begin{equation}
    \frac{\partial \mathcal{L}}{\partial \theta_i} = \frac{\partial \mathcal{L}}{\partial G} \frac{\partial G}{\partial \theta_i}.
    \label{eq:chain_rule}
    \end{equation}
    The gradient $\partial \mathcal{L} / \partial G$ is the same gate-matrix gradient computed through \ac{TN} contraction in Ref.~\cite{le2025riemannianquantum}; the chain rule reuses this gradient with the derivatives of the parameterized native gate. The product in \cref{eq:chain_rule} denotes contraction over the gate-matrix entries. The key difference is that the optimization proceeds directly in the (typically low-dimensional) Euclidean parameter space of the native gate, rather than being projected onto a Riemannian manifold. Because the resulting gates are already expressed in native form, no subsequent KAK synthesis is required, avoiding the associated blow-up in single- and two-qubit gate count.
    \item Conversely, single-qubit gates are optimized directly as generic unitaries via \ac{RO}. This allows the \ac{TN} environment
      tensor to be constructed efficiently around each single-qubit unitary, rather than separately for its constituent rotations.
      The optimized unitary can then be readily decomposed into elementary rotations using an Euler-angle decomposition, without introducing additional two-qubit gates.
\end{itemize}

\section{Analysis and Results}
\label{sec:analysis}

We study the \ac{TLFIM} and the
XXZ transverse-field Heisenberg model in one dimension, both on chains
of $N=50$ qubits with open boundary conditions. Neither model is
integrable in the parameter regimes considered. In the Ising model, the longitudinal field breaks the $\mathbb{Z}_2$ spin-flip symmetry. In the XXZ model, the transverse field breaks the $U(1)$ symmetry associated with spin rotations about the $z$-axis.
Because both models obey an area law in their ground states, \acp{MPS} represent them efficiently at $t=0$. During time evolution, entanglement often grows approximately linearly in time. Since an MPS with bond dimension $\chi$ can support an entanglement entropy of at most $\log \chi$, representing this growth requires $\chi \gtrsim e^{S(t)}$. The required bond dimension therefore grows exponentially in time. This exponential growth in classical memory and runtime motivates compressing the
time-evolution circuit for execution on quantum hardware rather than simulating it classically.

\subsection{Transverse-longitudinal field Ising model}
\label{sec:ising}

The \ac{TLFIM} is governed by the Hamiltonian
\begin{equation}
    \label{eq:ising_hamiltonian}
    H = \sum_{i} \left( J Z_i Z_{i+1} + h X_i + g Z_i \right).
\end{equation}
We consider a chain of $N=50$ sites. As the reference, we use the time-evolution
operator $U(t) = e^{-iHt}$ obtained from 20 repetitions of a fine fourth-order
Suzuki--Trotter decomposition based on the even--odd splitting of
\cref{eq:trotter_split_riemannian} below, which on hardware would require
603 $R_{ZZ}$ gates.

\subsubsection{Circuit ansätze and Trotter splitting}
\label{sec:ising_splitting}

Both methods are initialized from a second-order Trotterized brickwall circuit,
but they use different splittings of the Hamiltonian. The splitting determines
both the accuracy of the initial (unoptimized) circuit and the number of native
layers per Trotter step. For a fair comparison, we use the best available
splitting for each method.

\begin{figure}[t]
    \centering
    \includegraphics[width=\columnwidth]{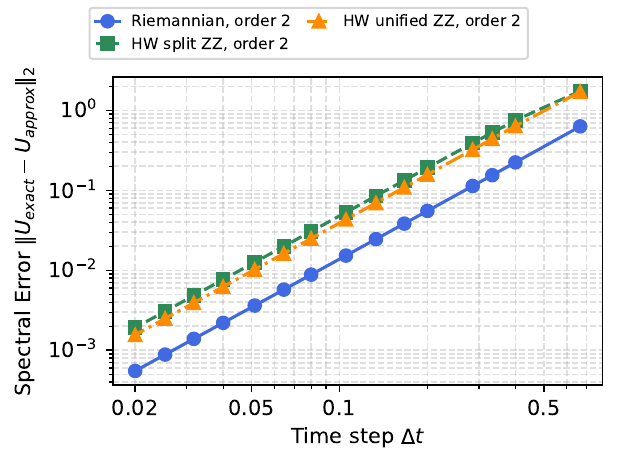}
    \caption{Estimated spectral error for different Trotter splittings as a
    function of the time step $\delta$ ($N=10$, $t=2$).
    Blue: splitting used for \ac{RO} (\cref{eq:trotter_split_riemannian}).
    Yellow and green: splittings available to \ac{PO}
    (\cref{eq:trotter_split_bad_hardware_aware} and
    \cref{eq:trotter_split_good_hardware_aware}, respectively). All lines share
    the same slope; the offset between them grows linearly with system size and
    interaction strength, reflecting the number of non-commuting terms introduced
    by each splitting.}
    \label{fig:trotter_split}
\end{figure}

For \ac{RO}, we use the even--odd splitting $H = H_{\text{even}} + H_{\text{odd}}$
of Ref.~\cite{le2025riemannianquantum}, in which the single-qubit field terms are
divided equally between the two halves:
\begin{subequations}
    \label{eq:trotter_split_riemannian}
\begin{align}
    H_{\text{even}} &= J\sum_{i \in \text{even}} Z_i Z_{i+1}
      + \frac{1}{2}\sum_i \left( h X_i + g Z_i \right), \\
    H_{\text{odd}} &= J\sum_{i \in \text{odd}} Z_i Z_{i+1}
      + \frac{1}{2}\sum_i \left( h X_i + g Z_i \right).
\end{align}
\end{subequations}
Of the splittings we tested, this one gives the smallest Trotter error
(\cref{fig:trotter_split}, blue). It is, however, unsuitable for \ac{PO}: each
bond gate combines the $ZZ$ interaction with the non-commuting $X$ field, so it is
a generic two-qubit unitary rather than an $R_{ZZ}$ rotation and would itself
require synthesis.

For \ac{PO}, we therefore keep the interaction and field terms in separate
layers, $H = H_{ZZ} + H_{\text{field}}$, with
\begin{subequations}
    \label{eq:trotter_split_good_hardware_aware}
\begin{align}
    H_{ZZ} &= J\sum_{i} Z_i Z_{i+1}, \\
    H_{\text{field}} &= \sum_{i} \left( h X_i + g Z_i \right).
\end{align}
\end{subequations}
For a time step $\delta$, every two-qubit gate is then a pure rotation
$e^{-i\delta J Z_i Z_{i+1}} = R_{ZZ}(2J\delta)$, with
$R_{ZZ}(\theta) = e^{-i\theta Z\otimes Z/2}$, and the field term contributes a
layer of single-qubit gates. An alternative is to place the field layer between
the even and odd bonds,
\begin{equation}
    \label{eq:trotter_split_bad_hardware_aware}
    H = H_{\text{even}, ZZ} + H_{\text{field}} + H_{\text{odd}, ZZ}.
\end{equation}
This saves one $R_{ZZ}$ layer in the second-order Trotter circuit (Appendix,
\cref{fig:circuit_field-middle}), but introduces more non-commuting terms per
step and hence a larger Trotter error (\cref{fig:trotter_split}, yellow vs.\
green). Because this error grows with system size, we use
\cref{eq:trotter_split_good_hardware_aware} throughout.

A natural question is whether \ac{RO} could also benefit from the lower $R_{ZZ}$
cost of the \ac{PO} splitting by using it as the initial circuit. It cannot:
during optimization, the $R_{ZZ}$ gates evolve into generic two-qubit unitaries,
each of which requires three $R_{ZZ}$ gates after synthesis, so the native gate
count would roughly triple.

\subsubsection{Degrees of freedom}
\label{sec:ising_dof}

The \ac{DOF} of a circuit depend on its gate count and on the manifold dimension
per gate: a generic two-qubit gate in $SU(4)$ has 15 real parameters, and a
single-qubit gate in $SU(2)$ has 3.

For a second-order Trotterization with $r$ repetitions, \ac{RO} produces $2r+1$
layers of $SU(4)$ gates: $r+1$ even-bond layers of $N/2$ gates each and $r$
odd-bond layers of $N/2-1$ gates each. The total gate count is
$G_{SU(4)} = (r+1)N/2 + r(N/2-1)$, giving
\begin{equation}
    D_{\text{RO}} = 15 \left[ (r+1)\frac{N}{2} + r\left(\frac{N}{2} - 1\right) \right].
\end{equation}

\begin{figure}[htbp]
    \centering
    \includegraphics[width=\linewidth]{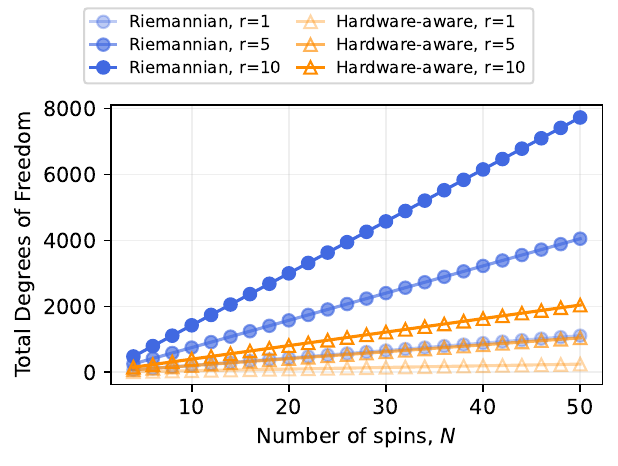}
    \caption{Total number of adjustable circuit parameters (degrees of freedom)
    versus the number of spins $N$ for Ising time-evolution circuits initialized
    from a second-order Trotterization. Circles represent \ac{RO}; triangles
    represent \ac{PO}. Shades indicate the number of Trotter repetitions $r$.
    Both parameter counts grow linearly with system size. At fixed $N$ and $r$,
    the hardware-aware circuits use fewer parameters, permitting more Trotter
    repetitions within the same parameter budget.}
    \label{fig:dof_comparison}
\end{figure}

For \ac{PO}, the same evolution time and order yields $2(r+1)$ interaction layers
($r+1$ even and $r+1$ odd layers of $R_{ZZ}$ gates, with one parameter each) and
$r$ layers of single-qubit rotations on all $N$ sites (three parameters each):
\begin{equation}
\begin{split}
    D_{\text{PO}} &= (r+1)\frac{N}{2} + (r+1)\left(\frac{N}{2}-1\right) + 3rN \\
    &= (r+1)(N-1) + 3rN.
\end{split}
\end{equation}
For $N=50$ and $r=5$, this gives $D_{\text{RO}} = 4050$ and $D_{\text{PO}} = 1044$.
The parameterized ansatz therefore reaches the same \ac{DOF} budget at a larger
number of Trotter repetitions, as shown in \cref{fig:dof_comparison}.

\subsubsection{Results}
\label{sec:ising_results}

\begin{figure}[t]
    \centering
    \includegraphics[width=\columnwidth]{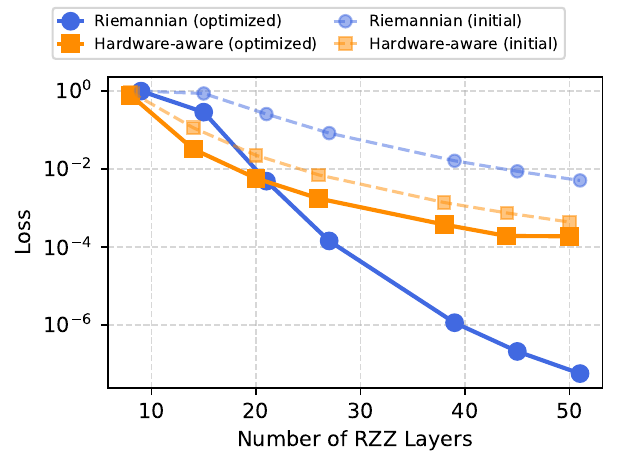}
    \caption{Loss as a function of native entangling-gate depth for the Ising
    model, $N=50$. \Ac{RO} is compared with \ac{PO}. Solid lines show results
    after optimization; dashed lines show the corresponding initial Trotter
    circuits before optimization. Lower loss indicates a more accurate
    approximation of the reference time evolution.}
    \label{fig:ising_results}
\end{figure}

\Cref{fig:ising_results} compares \ac{RO} and \ac{PO} as a function of native
entangling-gate depth. At low depth, \ac{PO} benefits from its Trotter-based
initialization, since more Trotter steps fit within a fixed gate budget. However,
optimizing its $R_{ZZ}$ parameters yields only marginal gains over the unoptimized
Trotterization. \ac{RO}, by contrast, overtakes \ac{PO} once the circuit exceeds
26 layers. This is consistent with Ref.~\cite{le2025riemannianquantum}, where the
Ising model is likewise the case in which \ac{RO} shows the clearest advantage.

We attribute the gap at higher depth to the limited expressivity of the native
two-qubit gate: each $R_{ZZ}$ gate is restricted to a one-dimensional segment of
the Weyl chamber (\cref{fig:weyl_chamber}), whereas each generic $SU(4)$ gate in
\ac{RO} can access the full chamber. The residual Weyl coordinates of
\cref{fig:weyl_ising}, which \ac{PO} neglects, therefore turn out to be essential
for high compression of the Ising propagator. Above 26 layers, \ac{PO} surpasses
\ac{RO} at only a single point in the tested range; we attribute this to the
sensitivity of \ac{RO} to the quality of its initialization, which leaves it prone
to local minima, an effect we revisit in \cref{sec:heisenberg}.

\subsection{The XXZ transverse-field Heisenberg model}
\label{sec:heisenberg}

The XXZ transverse-field Heisenberg model is given by
\begin{equation}
    \label{eq:heisenberg_hamiltonian}
H = \sum_{i} \left[ J(X_i X_{i+1} + Y_i Y_{i+1}) + J_z Z_i Z_{i+1} \right] + h \sum_{i} X_i,
\end{equation}
with $J$, $J_z$ the anisotropic exchange couplings and $h$ the
transverse-field strength.

\subsubsection{Native gate and Trotterization}

A continuous fSim gate is native to superconducting circuits with two
transmons coupled via a frequency-tunable coupler \cite{jiang2025concurrent},
and has also been proposed via a single set of flux pulses
\cite{foxen_2020_demonstrating}:

\begin{equation}
    \text{fSim}(\theta, \phi) = \begin{pmatrix} 
1 & 0 & 0 & 0 \\ 
0 & \cos\theta & -i\sin\theta & 0 \\ 
0 & -i\sin\theta & \cos\theta & 0 \\ 
0 & 0 & 0 & e^{-i\phi} 
\end{pmatrix}.
\end{equation}
As derived in Appendix \ref{app:heisenberg}, this gate is equivalent (up to local $Z$ rotations) to the XXZ interaction,
\begin{equation}
\text{fSim}(\theta,\phi) \approx \exp\left[-i t_1 (XX+YY) - i t_2 ZZ\right],
\end{equation}
with $\theta = 2t_1$, $\phi = 4t_2$. 

By combining a continuous fSim gate with an $R_{ZZ}$ pulse, any
arbitrary $SU(4)$ unitary can be synthesized (up to local gates) using only two entangling operations; we adopt this synthesis for the \acs{RO} circuits. For \acs{PO}, by contrast, we leverage the native XXZ (fSim) structure directly at initialization, splitting the Hamiltonian into even-bond, odd-bond, and transverse-field terms, with the constituent sub-Hamiltonians defined as:
\begin{subequations}
\begin{align}
H_{\text{even}} &= \sum_{i \in \text{even}} \left[ J(X_i X_{i+1} + Y_i Y_{i+1}) + J_z Z_i Z_{i+1} \right], \\
H_{\text{odd}} &= \sum_{i \in \text{odd}} \left[ J(X_i X_{i+1} + Y_i Y_{i+1}) + J_z Z_i Z_{i+1} \right], \\
H_{\text{field}} &= h \sum_{i} X_i.
\end{align}
\end{subequations}

The Trotterization for \acs{RO} mirrors the Ising case, distributing the transverse field terms symmetrically across both the even and odd bonds, where half of the local transverse field terms are absorbed directly into the two-body bond interactions:
\begin{align}
H_{\text{even/odd}}' &= \sum_{i \in \text{even/odd}} \left[ J(X_i X_{i+1} + Y_i Y_{i+1}) + J_z Z_i Z_{i+1} \right.\notag\\
&\qquad\qquad\left. + \frac{h}{2}(X_i + X_{i+1}) \right].
\end{align}

\begin{figure}[t]
    \centering
    \includegraphics[width=\columnwidth]{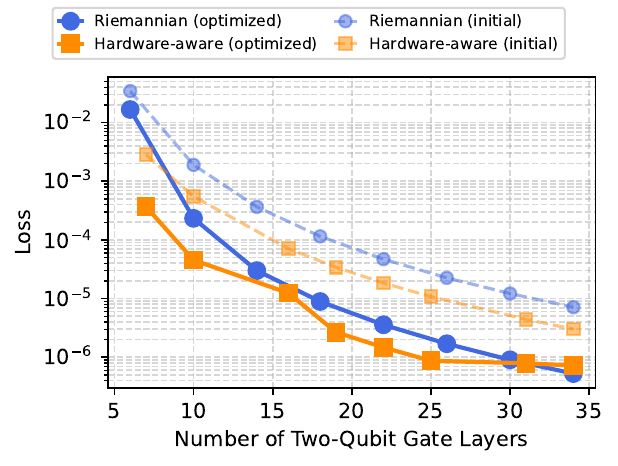}
    \caption{Loss as a function of native entangling gate depth for the Heisenberg
    XXZ model, $N=50$. Riemannian optimization (RO) is compared with hardware-aware
    parameterized optimization (PO). Each layer consists of native two-qubit gates: the \acs{RO} circuits alternate fSim and $R_{ZZ}$ layers, while the \acs{PO} circuits use only fSim layers. Solid lines show results after
    optimization; dashed lines show the corresponding initial Trotter circuits before optimization. Lower loss indicates a more accurate
    approximation to the reference time evolution.}
    \label{fig:heisenberg_results}
\end{figure}

The performance of the two compression methods is compared in
Fig.~\ref{fig:heisenberg_results}. In this Heisenberg setting, the depth reduction achieved by the hardware-aware approach is less pronounced than for the Ising model, since matching a generic two-qubit unitary now requires two native gates (an fSim and an $R_{ZZ}$) rather than three. For depths up to $\approx 35$ layers, \acs{PO} attains fidelities roughly half an order of magnitude better than \acs{RO} at equal depth. This advantage is inherited from both the initialization and the optimization: the hardware-aware
decomposition realizes each second-order Trotter step in fewer native layers, so at a fixed layer budget the \acs{PO} ansatz encodes a finer Trotter discretization and starts from a lower initial error.
Furthermore, the fSim gate is more expressive and entangling than the $R_{ZZ}$ gate, since it reaches an entire face of the Weyl chamber (Fig.~\ref{fig:weyl_chamber}) rather than a single line.

The \acs{RO} circuit (up to single-qubit transformations) depends heavily on the quality of its initial state and tends to become trapped in local minima, as reflected in the loss plateau seen in \cref{fig:riemannian_loss_history}. Whether this loss can be further reduced remains open; alternative optimization strategies, e.g.\ noise injection, may help escape these plateaus. The \acs{PO} circuit, by contrast, starts from a better initial state, and its gates are expressive enough to reach a higher fidelity early on. Beyond 35 layers, however, our hardware-aware training grows harder and longer, due to a higher number of tensor contractions, as explained in \cref{app:contraction_cost}) relative to \acs{RO}.

\color{black}

\section{Conclusion}
\label{sec:conclusion}
We compared two \ac{TN}-based compression strategies for Trotterized time-evolution circuits on a 50-qubit chain: standard \acs{RO}
over generic $SU(4)$ unitaries, and the hardware-aware \acs{PO} method that optimizes directly over the native two-qubit gate set.
The comparison was motivated by our observation that \acs{RO}-optimized gates already collapse into a predictable region of the Weyl chamber: near-$R_{ZZ}$
gates for the Ising model, and a local XXZ/fSim-like structure for the Heisenberg model. This structural regularity suggested that a shallower, hardware-native ansatz could recover most of \acs{RO}'s compression benefit
without the synthesis overhead of KAK decomposition.

The two methods perform very differently depending on the target Hamiltonian.
For the Ising case, \acs{RO} clearly outperforms \acs{PO}, particularly beyond $\sim$26 layers.  We attribute this gap to the expressivity of each two-qubit block: an \acs{RO} gate can be any element of $SU(4)$, realized by three $R_{ZZ}$ gates on the same pair of qubits, whereas in \acs{PO} each pair receives a single $R_{ZZ}$ per layer, restricted to one line of the Weyl chamber (\cref{fig:weyl_chamber}). Since a fixed-depth circuit can reach only a subset of the unitaries generated by its gates~\cite{barligea2026enablingliealgebraicclassicalsimulation}, this per-block restriction directly limits \acs{PO}. It matters even though the optimized \acs{RO} gates are close to $R_{ZZ}$ (\cref{fig:weyl_ising}), since their small $\alpha_y$ and $\alpha_z$ components cannot be produced by a single $R_{ZZ}$.
For the Heisenberg XXZ model, the outcome is reversed and, at first glance, surprising: \acs{PO} reaches a loss roughly half an order of magnitude lower than \acs{RO} for depths up to $\sim$35 layers, despite \acs{RO} being the more general method. We attribute this to two compounding effects:
(i) the hardware-aware Trotter splitting realizes each second-order step in fewer native layers, giving \acs{PO} a better-conditioned initial circuit at a fixed layer budget, and (ii) the fSim gate, generated by $XX+YY$ and $ZZ$ (up to single-qubit $Z$ phases), spans a two-parameter family covering an entire face of the Weyl chamber. Each \acs{PO} block therefore explores a substantially larger subgroup of $SU(4)$ than an $R_{ZZ}$ gate, which gives \acs{PO} enough expressivity to exploit its better starting point. \acs{RO}'s comparative weakness in this regime appears
to stem from sensitivity to initialization and susceptibility to local minima. Beyond approximately 35 layers, \acs{PO}'s advantage diminishes as optimization yields less improvement over the Trotter initialization. A possible explanation is correlated angle directions in finer Trotter decompositions. Establishing the mechanism requires further analysis of gradients and curvature, which we leave for future work. 

Beyond raw fidelity, \acs{PO} carries a substantial practical advantage in degrees of freedom: for $N=50$ and $r=5$ Trotter repetitions, \acs{PO} requires roughly a quarter of the parameters of \acs{RO} (1044 vs.\ 4050), which in turn permits more repetitions -- and thus finer Trotterizations -- within the same parameter
budget. However, this comes at a computational cost: \ac{TN} contraction scales as $O(\chi^3 D^3)$ for two-qubit gates and $O(\chi^3 D^2)$ for single-qubit gates, and since \acs{PO} retains Riemannian updates for its
single-qubit gates, its per-step cost remains higher than \acs{RO}'s, as detailed in \cref{app:contraction_cost}.

Taken together, these results indicate that neither method dominates uniformly: \acs{RO}'s generality pays off when the native entangling gate covers only a small region of the Weyl chamber, as $R_{ZZ}$ does in the Ising case, whereas \acs{PO}'s leaner, hardware-native parametrization is preferable when the native gate already covers the region relevant to the problem, as fSim does in the Heisenberg case, at least up to depths of $\sim$35 layers. More broadly, this suggests that exploiting the target hardware's native gate set -- and, in future work, its qubit connectivity -- is a promising direction for compressing circuits on both near-term and early fault-tolerant quantum computers.

This work focused on non-integrable models with no conserved observables of interest, and Trotterization itself breaks whatever symmetries the target Hamiltonian possesses. An open direction is to study how \acs{RO} and \acs{PO} compare when conserved quantities are relevant to the observables of interest, which likely calls for a different initialization strategy. Extending this comparison to fermionic systems is another natural avenue for future work. Under the Jordan--Wigner transformation, fermionic hopping and density--density interactions map directly onto the fSim gate, which forms the basis of efficient fermionic swap networks~\cite{kivlichan2019quantum}. This makes fermionic models a promising testbed for \acs{PO}.

\section*{Acknowledgments}
All tensor network operations were implemented using \texttt{jax.numpy} \cite{jax2018github}.
We thank Chinonso Onah, Roman Firt and Arne-Christian Voigt for valuable discussions over quantum native gates and support while running large scale experiments. 
The results, opinions and conclusions expressed in this publication are not necessarily those of Volkswagen AG.

\section*{Data availability}
The implementation used in this project, as well as the numerical results and a script to generate the figures in this publication, are available under  \url{https://github.com/ixaixim/weyl-circuit-compression}.

\bibliography{references}

\clearpage
\onecolumngrid
\appendix
\section{Ising Trotterization} 
\label{app:ising}
For the second-order Ising Trotterization with the splitting $H = H_{\text{ZZ}} + H_{\text{field}}$ of Eq.~\eqref{eq:trotter_split_good_hardware_aware}, the two-qubit depth scales as $d(r) = 2(r+1)$ when the interaction layers are placed outside (Fig.~\ref{fig:circuit_interaction-outside}) and as $d(r) = 2r$ when the field layers are placed outside (Fig.~\ref{fig:circuit_interaction-inside}). Placing the field between the even and odd bonds instead, as in Eq.~\eqref{eq:trotter_split_bad_hardware_aware}, gives $d(r) = 2r+1$ (Fig.~\ref{fig:circuit_field-middle}).
\tikzset{
    trotEven/.style={fill=evenColor},
    trotOdd/.style={fill=oddColor},
    trotLoc/.style={fill=gray!15},
}

\begin{figure}[htbp]
    \centering
    \begin{quantikz}[row sep=0.3cm, column sep=0.2cm]
        \lstick{$q_0$} & \gate[2, style={trotEven}]{R_{ZZ}(\Delta t/2)} & & \gate[style={trotLoc}]{U_{\mathrm{loc}}(\Delta t)} & \gate[2, style={trotEven}]{R_{ZZ}(\Delta t)} & & \gate[style={trotLoc}]{U_{\mathrm{loc}}(\Delta t)} & \gate[2, style={trotEven}]{R_{ZZ}(\Delta t/2)} & & \\
        \lstick{$q_1$} & & \gate[2, style={trotOdd}]{R_{ZZ}(\Delta t/2)} & \gate[style={trotLoc}]{U_{\mathrm{loc}}(\Delta t)} & & \gate[2, style={trotOdd}]{R_{ZZ}(\Delta t)} & \gate[style={trotLoc}]{U_{\mathrm{loc}}(\Delta t)} & & \gate[2, style={trotOdd}]{R_{ZZ}(\Delta t/2)} & \\
        \lstick{$q_2$} & \gate[2, style={trotEven}]{R_{ZZ}(\Delta t/2)} & & \gate[style={trotLoc}]{U_{\mathrm{loc}}(\Delta t)} & \gate[2, style={trotEven}]{R_{ZZ}(\Delta t)} & & \gate[style={trotLoc}]{U_{\mathrm{loc}}(\Delta t)} & \gate[2, style={trotEven}]{R_{ZZ}(\Delta t/2)} & & \\
        \lstick{$q_3$} & & & \gate[style={trotLoc}]{U_{\mathrm{loc}}(\Delta t)} & & & \gate[style={trotLoc}]{U_{\mathrm{loc}}(\Delta t)} & & &
    \end{quantikz}
    \caption{Two repetitions of the Interaction-Outside Trotter sequence for the splitting $H = H_{\text{ZZ}} + H_{\text{field}}$ [Eq.~\eqref{eq:trotter_split_good_hardware_aware}]. In the center, the interaction layers from both Trotter steps merge into a single $R_{ZZ}(\Delta t)$ operation.}
    \label{fig:circuit_interaction-outside}
\end{figure}
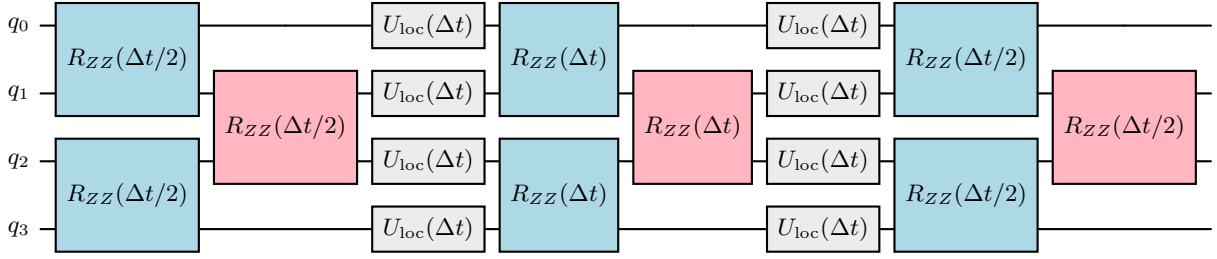

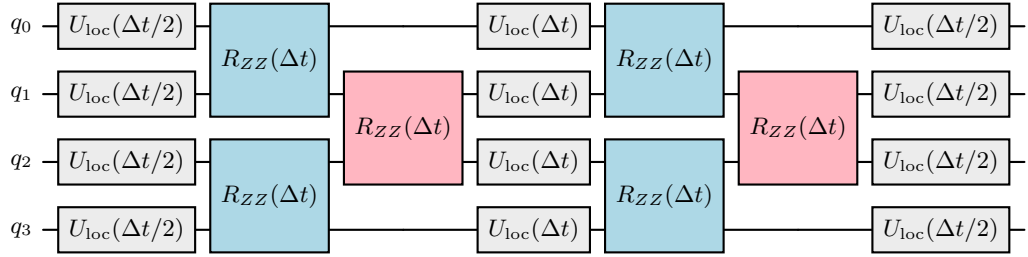
\begin{figure}[htbp]
    \centering
    \begin{quantikz}[row sep=0.3cm, column sep=0.2cm]
        \lstick{$q_0$} & \gate[style={trotLoc}]{U_{\mathrm{loc}}(\Delta t/2)} & \gate[2, style={trotEven}]{R_{ZZ}(\Delta t)} & & \gate[style={trotLoc}]{U_{\mathrm{loc}}(\Delta t)} & \gate[2, style={trotEven}]{R_{ZZ}(\Delta t)} & & \gate[style={trotLoc}]{U_{\mathrm{loc}}(\Delta t/2)} & \\
        \lstick{$q_1$} & \gate[style={trotLoc}]{U_{\mathrm{loc}}(\Delta t/2)} & & \gate[2, style={trotOdd}]{R_{ZZ}(\Delta t)} & \gate[style={trotLoc}]{U_{\mathrm{loc}}(\Delta t)} & & \gate[2, style={trotOdd}]{R_{ZZ}(\Delta t)} & \gate[style={trotLoc}]{U_{\mathrm{loc}}(\Delta t/2)} & \\
        \lstick{$q_2$} & \gate[style={trotLoc}]{U_{\mathrm{loc}}(\Delta t/2)} & \gate[2, style={trotEven}]{R_{ZZ}(\Delta t)} & & \gate[style={trotLoc}]{U_{\mathrm{loc}}(\Delta t)} & \gate[2, style={trotEven}]{R_{ZZ}(\Delta t)} & & \gate[style={trotLoc}]{U_{\mathrm{loc}}(\Delta t/2)} & \\
        \lstick{$q_3$} & \gate[style={trotLoc}]{U_{\mathrm{loc}}(\Delta t/2)} & & & \gate[style={trotLoc}]{U_{\mathrm{loc}}(\Delta t)} & & & \gate[style={trotLoc}]{U_{\mathrm{loc}}(\Delta t/2)} &
    \end{quantikz}
    \caption{Two repetitions of Interaction-Inside (Field-Outside) Trotterization for the same splitting [Eq.~\eqref{eq:trotter_split_good_hardware_aware}]. Note the merged central $U_{\mathrm{loc}}$ layer.}
    \label{fig:circuit_interaction-inside}
\end{figure}

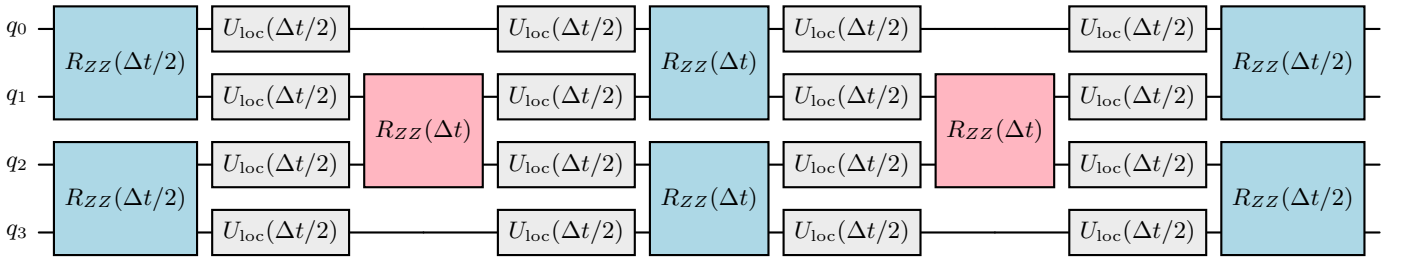
\begin{figure}[htbp]
    \centering
    \begin{quantikz}[row sep=0.3cm, column sep=0.2cm]
        \lstick{$q_0$} & \gate[2, style={trotEven}]{R_{ZZ}(\Delta t/2)} & \gate[style={trotLoc}]{U_{\mathrm{loc}}(\Delta t/2)} & & \gate[style={trotLoc}]{U_{\mathrm{loc}}(\Delta t/2)} & \gate[2, style={trotEven}]{R_{ZZ}(\Delta t)} & \gate[style={trotLoc}]{U_{\mathrm{loc}}(\Delta t/2)} & & \gate[style={trotLoc}]{U_{\mathrm{loc}}(\Delta t/2)} & \gate[2, style={trotEven}]{R_{ZZ}(\Delta t/2)} & \\
        \lstick{$q_1$} & & \gate[style={trotLoc}]{U_{\mathrm{loc}}(\Delta t/2)} & \gate[2, style={trotOdd}]{R_{ZZ}(\Delta t)} & \gate[style={trotLoc}]{U_{\mathrm{loc}}(\Delta t/2)} & & \gate[style={trotLoc}]{U_{\mathrm{loc}}(\Delta t/2)} & \gate[2, style={trotOdd}]{R_{ZZ}(\Delta t)} & \gate[style={trotLoc}]{U_{\mathrm{loc}}(\Delta t/2)} & & \\
        \lstick{$q_2$} & \gate[2, style={trotEven}]{R_{ZZ}(\Delta t/2)} & \gate[style={trotLoc}]{U_{\mathrm{loc}}(\Delta t/2)} & & \gate[style={trotLoc}]{U_{\mathrm{loc}}(\Delta t/2)} & \gate[2, style={trotEven}]{R_{ZZ}(\Delta t)} & \gate[style={trotLoc}]{U_{\mathrm{loc}}(\Delta t/2)} & & \gate[style={trotLoc}]{U_{\mathrm{loc}}(\Delta t/2)} & \gate[2, style={trotEven}]{R_{ZZ}(\Delta t/2)} & \\
        \lstick{$q_3$} & & \gate[style={trotLoc}]{U_{\mathrm{loc}}(\Delta t/2)} & & \gate[style={trotLoc}]{U_{\mathrm{loc}}(\Delta t/2)} & & \gate[style={trotLoc}]{U_{\mathrm{loc}}(\Delta t/2)} & & \gate[style={trotLoc}]{U_{\mathrm{loc}}(\Delta t/2)} & &
    \end{quantikz}
    \caption{Two repetitions of second-order Trotterization with the field placed between the even and odd bonds [Eq.~\eqref{eq:trotter_split_bad_hardware_aware}]. The even-bond layers of adjacent Trotter steps merge into a single $R_{ZZ}(\Delta t)$ operation, while the half-step field layers stay separated by the odd-bond layers.}
    \label{fig:circuit_field-middle}
\end{figure}

\section{The Heisenberg Model}
\label{app:heisenberg}
In the following, we prove: 
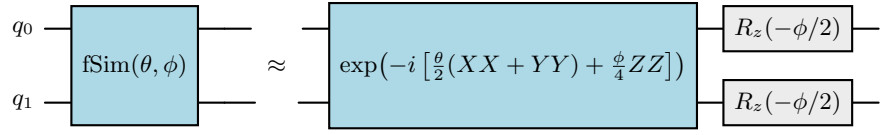
\begin{figure}[h]
    \centering
    \begin{quantikz}[row sep=0.3cm, column sep=0.35cm]
        \lstick{$q_0$} & \gate[2, style={fill=evenColor}]{\mathrm{fSim}(\theta, \phi)} & & \midstick[2, brackets=none]{$\approx$} & \gate[2, style={fill=evenColor}]{\exp\!\left(-i\left[\frac{\theta}{2}(XX+YY) + \frac{\phi}{4}ZZ\right]\right)} & \gate[style={fill=gray!15}]{R_z(-\phi/2)} & \\
        \lstick{$q_1$} & & & & & \gate[style={fill=gray!15}]{R_z(-\phi/2)} &
    \end{quantikz}
    \caption{Equivalence with explicit parameter mapping. To implement $\mathrm{fSim}(\theta, \phi)$, one evolves the XXZ Hamiltonian for $t_1 = \theta/2$ and $t_2 = \phi/4$, followed by single-qubit $Z$ rotations (grey).}
\end{figure}
\subsubsection{Proof}

To prove that the entangling layer is locally equivalent to the fSim gate, we begin by defining the rotation matrices $R_{XX}$, $R_{YY}$, and $R_{ZZ}$. These derivations rely on the property that for any Pauli tensor product $P$, $P^2 = I$. Using the identity $e^{i \alpha P} = \cos(\alpha) I + i \sin(\alpha) P$, we can express the rotations for an angle $\theta$.

First, we consider the matrices $XX$, $YY$, and $ZZ$:

$$XX = \begin{pmatrix} 0 & 0 & 0 & 1 \\ 0 & 0 & 1 & 0 \\ 0 & 1 & 0 & 0 \\ 1 & 0 & 0 & 0 \end{pmatrix}, \quad YY = \begin{pmatrix} 0 & 0 & 0 & -1 \\ 0 & 0 & 1 & 0 \\ 0 & 1 & 0 & 0 \\ -1 & 0 & 0 & 0 \end{pmatrix}, \quad ZZ = \begin{pmatrix} 1 & 0 & 0 & 0 \\ 0 & -1 & 0 & 0 \\ 0 & 0 & -1 & 0 \\ 0 & 0 & 0 & 1 \end{pmatrix}$$

The rotation matrices for these operators are given by:

$$R_{XX}(\theta) = e^{-i\frac{\theta}{2}XX} = \begin{pmatrix} \cos\frac{\theta}{2} & 0 & 0 & -i\sin\frac{\theta}{2} \\ 0 & \cos\frac{\theta}{2} & -i\sin\frac{\theta}{2} & 0 \\ 0 & -i\sin\frac{\theta}{2} & \cos\frac{\theta}{2} & 0 \\ -i\sin\frac{\theta}{2} & 0 & 0 & \cos\frac{\theta}{2} \end{pmatrix}$$

$$R_{YY}(\theta) = e^{-i\frac{\theta}{2}YY} = \begin{pmatrix} \cos\frac{\theta}{2} & 0 & 0 & i\sin\frac{\theta}{2} \\ 0 & \cos\frac{\theta}{2} & -i\sin\frac{\theta}{2} & 0 \\ 0 & -i\sin\frac{\theta}{2} & \cos\frac{\theta}{2} & 0 \\ i\sin\frac{\theta}{2} & 0 & 0 & \cos\frac{\theta}{2} \end{pmatrix}$$

For the combined $XX+YY$ interaction, the evolution $e^{-it_1(XX+YY)}$ yields:

$$e^{-it_1(XX+YY)} = \begin{pmatrix} 1 & 0 & 0 & 0 \\ 0 & \cos(2t_1) & -i\sin(2t_1) & 0 \\ 0 & -i\sin(2t_1) & \cos(2t_1) & 0 \\ 0 & 0 & 0 & 1 \end{pmatrix}$$

Next, we write the $ZZ$ evolution explicitly:

$$e^{-it_2 ZZ} = \begin{pmatrix} e^{-it_2} & 0 & 0 & 0 \\ 0 & e^{it_2} & 0 & 0 \\ 0 & 0 & e^{it_2} & 0 \\ 0 & 0 & 0 & e^{-it_2} \end{pmatrix}$$

Finally, combining these terms, we obtain the full entangling operation:

$$e^{-it_1(XX+YY)} e^{-it_2 ZZ} = \begin{pmatrix} e^{-it_2} & 0 & 0 & 0 \\ 0 & e^{it_2}\cos(2t_1) & -ie^{it_2}\sin(2t_1) & 0 \\ 0 & -ie^{it_2}\sin(2t_1) & e^{it_2}\cos(2t_1) & 0 \\ 0 & 0 & 0 & e^{-it_2} \end{pmatrix}$$

To transform the derived entangling operation $U_{\text{ent}} = e^{-it_1(XX+YY)} e^{-it_2 ZZ}$ into the standard \textbf{fSim} gate, we apply local single-qubit $Z$-rotations to both qubits. We define the local transformation $L$ as:

$$L = R_z(-2t_2) \otimes R_z(-2t_2)$$

Since $Z$-rotations commute with both $XX+YY$ and $ZZ$ interactions, this local operation can be applied either before or after $U_{\text{ent}}$.

\paragraph{Step-by-Step Verification}
Using the standard definition $R_z(\lambda) = \text{diag}(e^{-i\lambda/2}, e^{i\lambda/2})$, the local operator is:

$$R_z(-2t_2) \otimes R_z(-2t_2) = \text{diag}\left(e^{it_2} \cdot e^{it_2},\ e^{it_2} \cdot e^{-it_2},\ e^{-it_2} \cdot e^{it_2},\ e^{-it_2} \cdot e^{-it_2}\right) = \text{diag}\left(e^{i2t_2},\ 1,\ 1,\ e^{-i2t_2}\right)$$

Multiplying this local gate with the previously derived matrix $U_{\text{ent}}$:

$$L \cdot U_{\text{ent}} = \begin{pmatrix} 
e^{i2t_2}e^{-it_2} & 0 & 0 & 0 \\ 
0 & e^{it_2}\cos(2t_1) & -ie^{it_2}\sin(2t_1) & 0 \\ 
0 & -ie^{it_2}\sin(2t_1) & e^{it_2}\cos(2t_1) & 0 \\ 
0 & 0 & 0 & e^{-i2t_2}e^{-it_2} 
\end{pmatrix} = \begin{pmatrix} 
e^{it_2} & 0 & 0 & 0 \\ 
0 & e^{it_2}\cos(2t_1) & -ie^{it_2}\sin(2t_1) & 0 \\ 
0 & -ie^{it_2}\sin(2t_1) & e^{it_2}\cos(2t_1) & 0 \\ 
0 & 0 & 0 & e^{-i3t_2} 
\end{pmatrix}$$

By factoring out the global phase $e^{it_2}$, we obtain:

$$e^{-it_2} (L \cdot U_{\text{ent}}) = \begin{pmatrix} 
1 & 0 & 0 & 0 \\ 
0 & \cos(2t_1) & -i\sin(2t_1) & 0 \\ 
0 & -i\sin(2t_1) & \cos(2t_1) & 0 \\ 
0 & 0 & 0 & e^{-i4t_2} 
\end{pmatrix}$$

This result is identical to the standard \textbf{fSim} gate matrix:

$$\text{fSim}(\theta, \phi) = \begin{pmatrix} 
1 & 0 & 0 & 0 \\ 
0 & \cos\theta & -i\sin\theta & 0 \\ 
0 & -i\sin\theta & \cos\theta & 0 \\ 
0 & 0 & 0 & e^{-i\phi} 
\end{pmatrix}$$

where the parameters are mapped as:

$$\boxed{\theta = 2t_1, \quad \phi = 4t_2}$$

In practical hardware implementations, such as superconducting circuits, these single-qubit $Z$ rotations are implemented as virtual phase updates at zero cost, making the entangling layer locally equivalent to the fSim gate.

\clearpage
\section{Optimization}
\label{sec:optimization}
\begin{figure}[!h]
    \centering
    \includegraphics[width=0.8\textwidth]{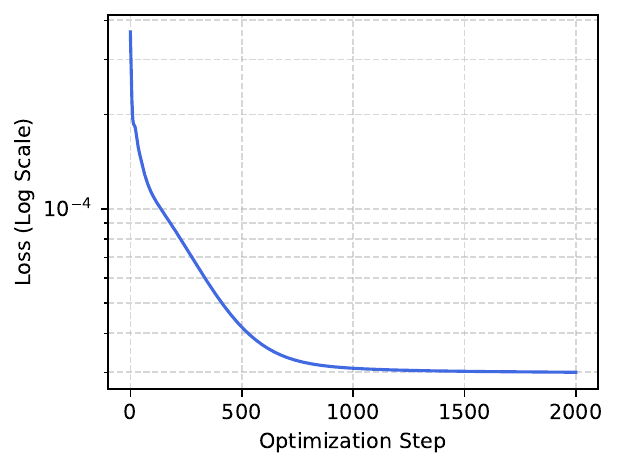}
    \caption{Riemannian Adam Loss History for a XXZ Heisenberg system with 50 sites and initialization to 5 repetitions of second-order Trotterization.}
    \label{fig:riemannian_loss_history}
\end{figure}

Fig.~\ref{fig:riemannian_loss_history} shows the loss history of Riemannian Adam for the XXZ Heisenberg model with 50 sites, initialized with 5 repetitions of second-order Trotterization. After a rapid initial decrease, the loss levels off into a plateau. Such plateaus are indicative of local minima of the Riemannian loss function.
\clearpage

\section{Contraction cost}
\label{app:contraction_cost}

In this appendix we detail the computational cost of contracting the two elementary building blocks of a Trotter layer -- single- and two-qubit gates -- into the \acs{TN} used to evaluate the loss function and its gradient during optimization. Namely, we are showing the cost of updating a left environment during a sweep with the tensor contraction presented in \cite{le2025riemannianquantum}. Because a \acs{PO} circuit interleaves single-qubit gates between the two-qubit entangling gates, while an \acs{RO} circuit is built exclusively from two-qubit gates, the \acs{TN} underlying a \acs{PO} layer
contains additional tensors to contract relative to an \acs{RO} layer of the same width, as illustrated in Fig.~\ref{fig:trotter_comparison}. Throughout, $\chi$ denotes the bond dimension of the boundary environment $L$ (and of the \acs{MPO} tensors $A_1,A_2,B_1,B_2$ below), while $d$ denotes the local dimension of a gate leg
($d=2$ for a single qubit); 

\subsection{Two-qubit gate}

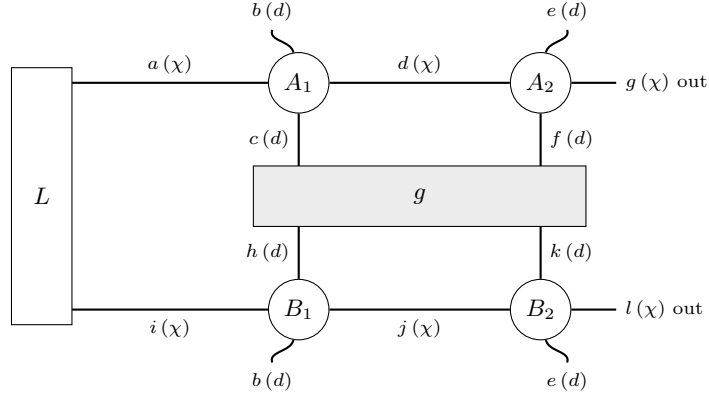
\begin{figure}[t]
\centering
\begin{tikzpicture}[
  tens/.style={draw, circle, fill=white, minimum size=0.8cm, inner sep=0pt, font=\small},
  Lbox/.style={draw, rectangle, fill=white, minimum width=0.8cm, minimum height=3.4cm, font=\small},
  gate/.style={draw, rectangle, minimum width=4.4cm, minimum height=0.8cm, fill=gray!15, font=\small},
  bond/.style={thick},
  lbl/.style={font=\scriptsize}
]
  \node[Lbox] (L)  at (-2,1.5)  {$L$};
  \node[tens] (A1) at (1.4,3)   {$A_1$};
  \node[tens] (A2) at (4.6,3)   {$A_2$};
  \node[gate] (g)  at (3.0,1.5) {$g$};
  \node[tens] (B1) at (1.4,0)   {$B_1$};
  \node[tens] (B2) at (4.6,0)   {$B_2$};

  \draw[bond] (-1.6,3) -- (A1) node[lbl, midway, above] {$a\,(\chi)$};
  \draw[bond] (-1.6,0) -- (B1) node[lbl, midway, below] {$i\,(\chi)$};

  \draw[bond] (A1) -- (A2) node[lbl, midway, above] {$d\,(\chi)$};
  \draw[bond] (B1) -- (B2) node[lbl, midway, below] {$j\,(\chi)$};

  \draw[bond] (A1) -- (1.4,1.9) node[lbl, midway, left]  {$c\,(d)$};
  \draw[bond] (A2) -- (4.6,1.9) node[lbl, midway, right] {$f\,(d)$};
  \draw[bond] (B1) -- (1.4,1.1) node[lbl, midway, left]  {$h\,(d)$};
  \draw[bond] (B2) -- (4.6,1.1) node[lbl, midway, right] {$k\,(d)$};

  \draw[bond] (A2) -- ++(1,0) node[lbl, right] {$g\,(\chi)$ out};
  \draw[bond] (B2) -- ++(1,0) node[lbl, right] {$l\,(\chi)$ out};

  \draw[bond] (A1) to[out=100,in=250] ++(-0.35,0.7)  node[lbl, above] {$b\,(d)$};
  \draw[bond] (B1) to[out=260,in=110] ++(-0.35,-0.7) node[lbl, below] {$b\,(d)$};
  \draw[bond] (A2) to[out=80,in=290]  ++(0.35,0.7)   node[lbl, above] {$e\,(d)$};
  \draw[bond] (B2) to[out=280,in=70]  ++(0.35,-0.7)  node[lbl, below] {$e\,(d)$};
\end{tikzpicture}
\caption{Contraction scheme for a two-qubit gate $g$, contracted with a left
environment $L$ and the corresponding MPO tensors $A_1,A_2,B_1,B_2$ of the
sites on which the gate acts.}
\label{fig:two_qubit_network}
\end{figure}

Figure~\ref{fig:two_qubit_network} shows the \acs{TN} encountered when contracting a two-qubit gate $g$, acting on sites $i,i+1$, against the accumulated left environment $L$ and the four neighboring \acs{MPO} tensors $A_1,A_2$ (upper/bra environment) and $B_1,B_2$ (lower/ket environment). The environment $L$ carries the two bond-dimension-$\chi$ legs $a,i$ inherited
from the previously contracted sites. 

We contract this network sequentially into an updated environment $L'$,
absorbing one tensor at a time in the order $A_1 \to B_1 \to g \to A_2 \to
B_2$. Table~\ref{tab:two_qubit_cost}
lists, for each step, the indices summed over and the resulting cost,
computed as the product of the dimensions of every leg touched (summed or
left open) in that step.

\begin{table}[h]
\centering
\begin{tabular}{c c c c}
\hline\hline
Step & Absorb & Contract & Cost \\
\hline
1 & $A_1$ & $a$     & $\chi^3 d^2$ \\
2 & $B_1$ & $i,b$   & $\chi^3 d^3$ \\
3 & $g$   & $c,h$   & $\chi^2 d^4$ \\
4 & $A_2$ & $d,f$   & $\chi^3 d^3$ \\
5 & $B_2$ & $j,k,e$ & $\chi^3 d^2$ \\
\hline\hline
\end{tabular}
\caption{Cost of each step in the sequential contraction of a two-qubit
gate.}
\label{tab:two_qubit_cost}
\end{table}

Since the gate leg dimension is small compared to the bond dimensions used
in this work ($d=2 \ll \chi$), the two steps absorbing $B_1$ and $A_2$
dominate, and contracting a single two-qubit gate into the environment
scales as
\begin{equation}
O(\chi^3 d^3).
\end{equation}

\subsection{Single-qubit gate}

\begin{figure}[t]
\centering
\begin{tikzpicture}[
  tens/.style={draw, circle, fill=white, minimum size=0.8cm, inner sep=0pt, font=\small},
  Lbox/.style={draw, rectangle, fill=white, minimum width=0.8cm, minimum height=3.4cm, font=\small},
  gate/.style={draw, rectangle, minimum width=0.8cm, minimum height=0.8cm, fill=gray!15, font=\small},
  bond/.style={thick},
  lbl/.style={font=\scriptsize}
]
  \node[Lbox] (L)  at (-2,1.5)  {$L$};
  \node[tens] (A1) at (1.4,3)   {$A_1$};
  \node[gate] (g)  at (1.4,1.5) {$g$};
  \node[tens] (B1) at (1.4,0)   {$B_1$};

  \draw[bond] (-1.6,3) -- (A1) node[lbl, midway, above] {$a\,(\chi)$};
  \draw[bond] (-1.6,0) -- (B1) node[lbl, midway, below] {$i\,(\chi)$};

  \draw[bond] (A1) -- (1.4,1.9) node[lbl, midway, left] {$c\,(d)$};
  \draw[bond] (B1) -- (1.4,1.1) node[lbl, midway, left] {$h\,(d)$};

  \draw[bond] (A1) -- ++(1,0) node[lbl, right] {$d\,(\chi)$ out};
  \draw[bond] (B1) -- ++(1,0) node[lbl, right] {$j\,(\chi)$ out};

  \draw[bond] (A1) to[out=100,in=250] ++(-0.35,0.7)  node[lbl, above] {$b\,(d)$};
  \draw[bond] (B1) to[out=260,in=110] ++(-0.35,-0.7) node[lbl, below] {$b\,(d)$};
\end{tikzpicture}
\caption{Contraction scheme for a single-qubit gate $g$, contracted with a
left environment $L$, an upper tensor $A_1$, and a lower tensor $B_1$ from
the respective upper and lower environments.}
\label{fig:one_qubit_network}
\end{figure}
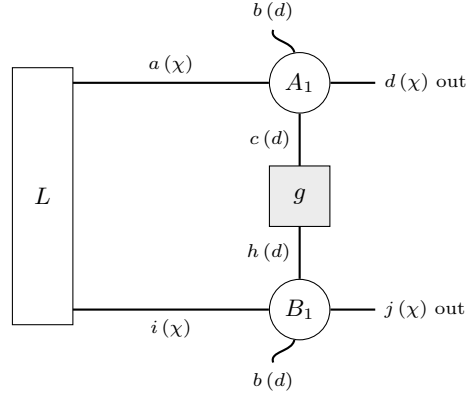

The corresponding network for a single-qubit gate is shown in
Fig.~\ref{fig:one_qubit_network}: the gate $g$ acts on a single upper tensor
$A_1$ and lower tensor $B_1$, each connected to $L$ by a $\chi$-bond ($a$,
$i$) and to each other by a direct $d$-dimensional  bond $b$.

Absorbing $A_1$, then the gate, then $B_1$ (Table~\ref{tab:one_qubit_cost}) updates the environment to $L'$.

\begin{table}[h]
\centering
\begin{tabular}{c c c c}
\hline\hline
Step & Absorb & Contract & Cost \\
\hline
1 & $A_1$ & $a$     & $\chi^3 d^2$ \\
2 & $g$   & $c$     & $\chi^2 d^3$ \\
3 & $B_1$ & $i,b,h$ & $\chi^3 d^2$ \\
\hline\hline
\end{tabular}
\caption{Cost of each step in the sequential contraction of a single-qubit
gate.}
\label{tab:one_qubit_cost}
\end{table}

Here the first and last steps dominate, so contracting a single-qubit gate
into the environment scales as
\begin{equation}
O(\chi^3 d^2).
\end{equation}

These per-gate costs directly account for the runtime gap discussed in
Sec.~IV: on top of the $O(\chi^3d^3)$ cost of the two-qubit gates common to
both methods, each \acs{PO} Trotter layer additionally requires the
$O(\chi^3d^2)$ contraction of $N$ single-qubit gates, retained as
Riemannian updates, which \acs{RO} does not incur.
\end{document}